\documentclass[a4paper, 10pt, notitlepage]{article}
\usepackage{amsmath,amssymb,enumerate,subfig,comment,epsfig,epstopdf,morefloats, authblk}
\usepackage{pgfplots,tikz,float,graphicx,caption,accents}
\usepackage{setspace,titling,natbib}
\usepackage{url}
\usepackage[left=1in, right=1in, top=1in, bottom=1in]{geometry}

\pretitle{\begin{center}\large\bfseries}
\posttitle{\par\end{center}\vskip 0.5em}
\preauthor{\begin{center}\large\ttfamily}
\postauthor{\end{center}}
\predate{\par\normalsize\centering}
\postdate{\par\small\centering}

\title{Modelling pulsatile blood flow through surgically coupled microvascular anastomoses.}
\author{M.~T.~Gallagher}
\date{\today}

\affil{School of Mathematics, University of Birmingham,Birmingham. B15 2TT. UK}

\providecommand{\bra}[1]{\left(#1\right)}
\providecommand{\Bra}[1]{\left\lbrace #1\right\rbrace}
\providecommand{\sqbra}[1]{\left[#1\right]}
\providecommand{\abs}[1]{\left|#1\right|}
\providecommand{\uv}[1]{\hat{\mathbf{#1}}} 
\newcommand{\D}{\mathcal{D}}
\newcommand{\R}{\mathbb{R}}
\providecommand{\ezb}{e^{-z^2/\beta^2}}
\providecommand{\ezbp}[1]{e^{-{#1}z^2/\beta^2}}
\providecommand{\Jor}{J_0\bra{i^{3/2}\alpha r}}
\providecommand{\Jo}{J_0\bra{i^{3/2}\alpha}}
\providecommand{\Jir}{J_1\bra{i^{3/2}\alpha r}}
\providecommand{\Ji}{J_1\bra{i^{3/2}\alpha}}
\providecommand{\Jbar}{\bar{\mathrm{J}}}
\providecommand{\pdiff}[2]{\frac{\partial{#1}}{\partial{#2}}}

\begin{document}

\maketitle

\begin{abstract}

The significant rates of post-operative arterial thrombus formation following the sutured anastomosis (surgical connection) of vessels in microvascular reconstruction has kindled the development of engineered solutions in the form of surgical coupling devices. These devices reduce thrombosis rates, which recent numerical studies suggest to be due to decreases in shear strain rate across the anastomosis site when compared to sutured vessels. In this work we develop and analytically solve the first mathematical model for the problem of microvascular anastomosis, leading to the discovery that the rates of thrombosis using these surgical coupling devices could be further reduced by decreasing the gradient of the wall deformation at the join.

\end{abstract}

\section{Introduction}
\label{sec:intro}

The ability to create a well structured anastomosis, the surgical connection between two vessels, is essential in a whole range of medical applications, particularly in the case of microvascular reconstruction. It is well known, for example, that the survival of tissue post-breast reconstructive surgery is highly dependent on the ability of the connected vasculature to flow freely \citep{koul2013}, with a leading cause of failure being that of thrombosis \citep{kroll1996}. This has led to the recent series of papers by Wain \textit{et al.} (\citet{wain2014}, \citet{wain2016}, \citet{wain2018})  who performed the first detailed numerical investigations into the flow properties through microvascular anastomoses.

Currently the preferred method for vascular anastomoses involves the end-to-end suturing of vessels \citep{carrel1902}. However, for smaller vessels in the microvasculature, even when executed by the most highly trained of surgeons, this technique can result in thrombosis rates of around $ 3.3\% $ \citep{yap2006} and can be very time consuming. As a result of this, there has engineered solutions have been developed in the form of coupling devices such as the UNILINK coupler \citep{ostrup1986}, see figure~\ref{fig:sketch}a, which has been shown to result in thrombosis rates as low as $ 0.6\% $ \citep{jandali2010}, with several sources reporting the decrease in both the failure rates and the time required for anastomoses when compared to the use of sutures \citet{de1996}. The UNILINK device was designed for venous anastomoses and as such is not recommended for arterial anastomoses, although this has not prevented the developing use of such devices in breast reconstruction \citep{spector2006}, and free tissue transfer \citep{grewal2012}. This push towards the use of surgical coupling devices on small vessels for \textit{microvascular arterial} anastomoses means that it is more important than ever that we fully understand the detailed properties of how the flow is affected by this procedure. 

The key influence on the rates of thrombosis is the shear rate on the wall of the vessel, with increases in such having been shown to lead to increases in thrombus formation \citep{sakariassen2015} due to increased platelet activation and aggregation at the site of anastomosis \citep{murray1926}. In addition to increases in the magnitude of shear rate, it has been shown by \citet{nesbitt2009} that shear gradients play a significant role in the rate and size of platelet aggregation, and thus the potential formation of thrombus. 

With this in mind, recent studies have been investigating in detail the effect of both surgical coupling and suturing on the flow through small vessels approximately $ 1 $~mm in radius, with particular attention being paid to the shear rate across the anastomosis. These studies, performed numerically with computational fluid dynamics (CFD) packages such as ANSYS-CFX, have aimed to develop increasingly realistic models of microvascular anastomoses, beginning by investigating the flow due to a constant pressure gradient \citep{wain2014}, followed by detailed studies into the effect of suture placement \citep{wain2016}, and, most recently, considering the effects of including flows driven by pulsatile pressure gradients \citep{wain2018}. To complement these studies, we develop here a mathematical model for the problem, which, once solved analytically, will allow the more detailed analysis of how the shear rate is influenced by the inclusion of a microvascular anastomosis.

The use of mathematical models has been fundamental to understanding all aspects of fluid dynamics, not-least those problems relating to the flow of fluid through arteries. Of particular interest for the present work is the first analytical expression for the velocity of a viscous incompressible fluid through an axisymmetric pipe, derived by \citet{womersley1955method}. In this study we will utilise the perturbation techniques of \citet{van1964} to consider the asymptotic correction terms to the Womersley flow profile due to a microvascular anastomosis, simplifying the problem by noting that we can consider sections of vessel which have a radial length scale that is much smaller than the axial length scale. The use of such simplifications, as well as other mathematical techniques, have been widely used to great success for related problems in blood flow; for a detailed explanation of the subject see \citet{pedley1980}. These techniques will enable significant analytical progress to be made in understanding the changes in shear rate across the site of anastomosis, and thus will bring new insight into the ways in which rates of thrombosis may be further reduced.

In this work, we develop the first mathematical model for coupled microvascular anastomosis in small vessels. By modelling the anastomosis as a small axisymmetric perturbation to a pipe with rigid walls, and considering an extruding bump as a model of wall deformation due to a coupling device, and an intruding bump as a first approximation to the deformation due to sutures, we investigate in detail the asymptotic correction to the flow profile driven by a pulsatile pressure gradient. We begin by deriving the governing equations for the problem in \S\ref{sec:eqns}, followed by the construction of the associated analytical solutions in \S\ref{sec:sol}. The solutions are discussed in \S\ref{sec:disc}, with particular attention paid to the physiological importance of the results.


\section{Equations of flow}
\label{sec:eqns}

\begin{figure}
	\centering
	\includegraphics[width=\textwidth]{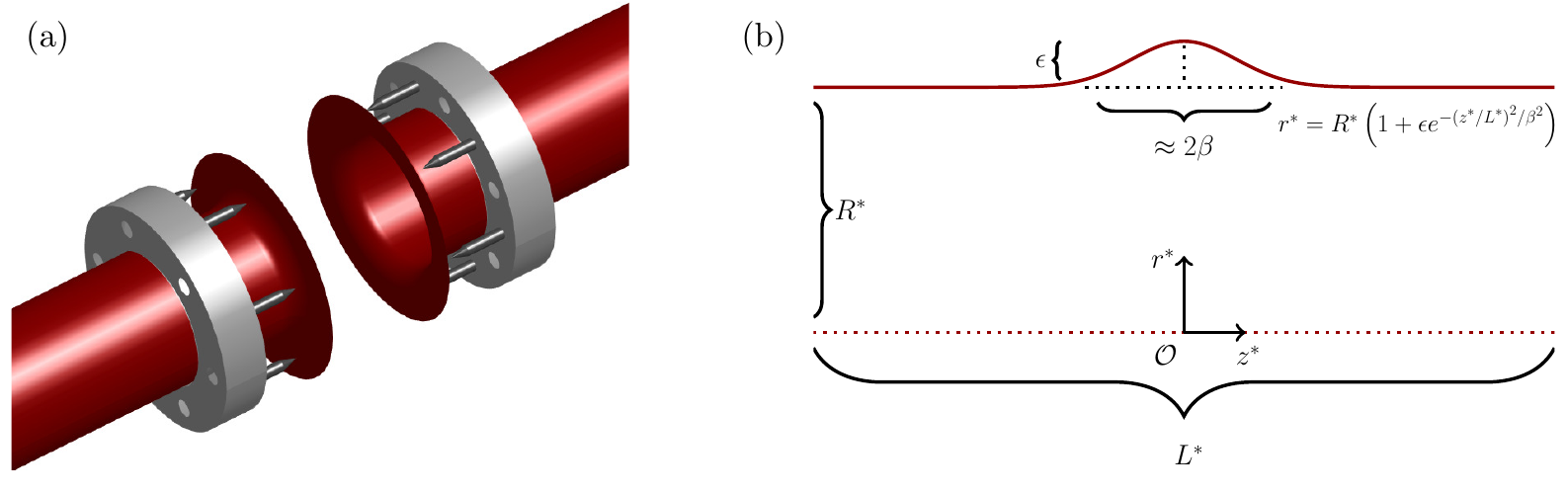}
	\caption{Definition sketches showing (a) the surgical coupler (redrawn from \citet{wain2014}) and (b) the modelled wall geometry post-surgical coupling.}
	\label{fig:sketch}
\end{figure}

We consider the flow of an incompressible Newtonian fluid driven by a pulsatile pressure gradient through a pipe of length $ L^* $ and radius $ R^* $, where $ \epsilon = R^* / L^* \ll 1 $. We model the site of anastomosis by an $ O\bra{\epsilon} $ axisymmetric deformation to the pipe wall centred about $ z^* = 0 $. Here, $ \bra{r^*,z^*} $ are the radial and axial components of a cylindrical polar coordinate system, with $ z^* $ pointing in the direction of flow, $ r^* $ pointing outwardly from the centre of the pipe, with time $ t^*  > 0 $, and $ ^* $ denoting dimensionful quantities. A definition sketch of the geometry is included in figure \ref{fig:sketch}b. We study the flow driven by an oscillatory pressure gradient of the form
\begin{equation}
	p_{0,z^*}^{*\pm}\bra{r^*,z^*,t^*} = P_0^* + P_1^* e^{i2\pi\omega^* t^*},
\end{equation}
for some constants $ P_0^* $ and $ P_1^* $, with $ \omega^* $ being the frequency of oscillation. Here, and throughout this paper, we use subscript notation to denote partial differentiation. We note that, when discussing the results of this model, we will consider only the real components of fluid pressure and velocity, however for algebraic simplicity we will not include this in our notation. We denote the axisymmetric fluid velocity by $ {\mathbf{u}^*\bra{r^*,z^*,t^*} = \bra{u^*\bra{r^*,z^*,t^*},0,w^*\bra{r^*,z^*,t^*}}} $, and introduce the dimensionless variables
	\begin{eqnarray}
		r^* &=& R^* r,\enspace z^* = \epsilon^{-1}R^{*} z,\enspace t^* = \bra{2\pi\omega}^{*-1}t,\nonumber\\
		u^* &=& \epsilon \omega^* R^* u,\enspace w^* = \omega^* R^* w,\enspace p^* = \epsilon^{-1}\omega^*\mu^*~ p,
		\label{eqn:dimensionless}
	\end{eqnarray}
where $ \mu^* $ is the fluid viscosity. The flow is characterised by the dimensionless Womersley number \citep{womersley1955method}, a measure of the relative importance of the pulsatile flow frequency to viscous effects, namely
\begin{equation}
	\alpha = R^*\sqrt{\frac{\rho^*\omega^*}{\mu^*}},
	\label{eqn:wom}
\end{equation}
with $ \rho^* $ the fluid density. We model the site of anastomosis by an axisymmetric Gaussian bump, with the boundary location given by
\begin{equation}
	r = h^{\pm}\bra{z} = 1 \pm \epsilon e^{-z^2/\beta^2},
	 \label{eqn:wall_dim}
\end{equation}
where $ \beta $ is the width parameter for the deformation.  Here, the case of a protruding bump ($h^{+}$ in~(\ref{eqn:wall_dim})) is a model for an arterial coupling device, while the intruding bump ($h^{-}$ in~(\ref{eqn:wall_dim})) is a first approximation towards modelling sutures. We write the velocity of the fluid under axisymmetry as
\begin{equation}
	\mathbf{u}^{\pm}\bra{r,z,t} = \epsilon u^\pm\bra{r,z,t}\uv{r} + w^\pm\bra{r,z,t}\uv{z},\quad \bra{r,z}\in\D^\pm\bra{t},\ t > 0,
	\label{eqn:gov_vel}
\end{equation}
where $ \uv{r} $ and $ \uv{z} $ are unit vectors in the radial and axial directions respectively, and the domain occupied by the fluid at time $ t $ is then given by
\begin{equation}
	\D^{\pm}\bra{t} = \Bra{\bra{r,\theta,z}\in\R^3 : r\in\left[0,h^{\pm}\bra{z}\right),\ \theta\in\left[0,2\pi\right), z\in\bra{-\infty,\infty}},\quad t\in\left[0,\infty\right).
\end{equation}
In what follows we will drop the superscript $ ^{\pm} $ for ease of notation. 

The flow is governed by the Navier-Stokes equations, which are given, in terms of the dimensionless variables~(\ref{eqn:dimensionless}), in component form, as
\begin{eqnarray}
	\epsilon^2\alpha^2 \sqbra{u_t + \epsilon\bra{u u_r + w u_z }} &=& -p_r + \epsilon^2\sqbra{\frac{1}{r}\bra{ru_r}_r - \frac{1}{r^2}u} + \epsilon^4 u_{zz},\label{eqn:gov_u}\\
	\alpha^2 \sqbra{w_t + \epsilon\bra{u w_r + w w_z }} &=& -p_z + \frac{1}{r}\bra{rw_r}_r + \epsilon^2 w_{zz},\label{eqn:gov_w}
\end{eqnarray}
together with conservation of mass,
\begin{equation}
	\frac{1}{r}\bra{ru}_r + w_z = 0,
	\label{eqn:gov_cont}
\end{equation}
each evaluated at all points in the domain $ \bra{r,z}\in\D\bra{t} $, $ t > 0 $, and completed by the no-slip, and no-penetration boundary conditions, namely
\begin{eqnarray}
	\mathbf{u}\cdot\uv{t}&= 0\quad\mbox{on}\quad r = h\bra{z},\ z\in\bra{-\infty,\infty},\ t > 0;\label{eqn:gov_noslip_full}\\
	\mathbf{u}\cdot\uv{n}&= 0\quad\mbox{on}\quad r = h\bra{z},\ z\in\bra{-\infty,\infty},\ t > 0;\label{eqn:gov_nopen_full}
\end{eqnarray}
and the axisymmetry conditions,
\begin{equation}
	u = 0,\quad w_r = 0\quad\mbox{on}\quad r = 0,\ z\in\bra{-\infty,\infty},\ t > 0.
	\label{eqn:gov_axisym}
\end{equation}
Here, the tangent vector in the direction of flow, $\uv{t}$, and outwardly pointing normal vector, $\uv{n}$, are given, after expanding in powers of $\epsilon$, as
\begin{eqnarray}
	\uv{t} &=& \bra{\mp\epsilon\frac{2z}{\beta^2}\ezb \pm \epsilon^3\frac{4z^3}{\beta^6}\ezbp{3} + O\bra{\epsilon^5}}\uv{r}\nonumber\\
	&&\qquad\qquad+ \bra{1-\epsilon^2\frac{2z^2}{\beta^4}\ezbp{2} + O\bra{\epsilon^4}}\uv{z},\label{eqn:gov_t}\\
	\uv{n} &=& \bra{1-\epsilon^2\frac{2z^2}{\beta^4}\ezbp{2} + O\bra{\epsilon^4}}\uv{r}\nonumber\\
	&&\qquad\qquad + \bra{\pm\epsilon\frac{2z}{\beta^2}\ezb \mp \epsilon^3\frac{4z^3}{\beta^6}\ezbp{3}+O\bra{\epsilon^5}}\uv{z}.\label{eqn:gov_n}
\end{eqnarray}
We expand the boundary conditions on the wall,~(\ref{eqn:gov_noslip_full}) and~(\ref{eqn:gov_nopen_full}), about $r = 1$, utilising the $O\bra{\epsilon}$ nature of the boundary perturbation, resulting in the simplified the no-slip condition
\begin{equation}
	w\bra{1,z,t} \pm \epsilon\ezb w_r\bra{1,z,t}  + O\bra{\epsilon^2} = 0,
	\label{eqn:gov_noslip}
\end{equation}
	and equivalent no-penetration condition 
\begin{eqnarray}
	u\bra{1,z,t} &\pm& \frac{2z}{\beta^2}\ezb w\bra{1,z,t}\nonumber\\
	&&\quad + \epsilon\bra{\frac{2z}{\beta^2}\ezbp{2} w_r\bra{1,z,t} \pm \ezb u_r\bra{1,z,t}} + O\bra{\epsilon^2} = 0,
	\label{eqn:gov_nopen}
\end{eqnarray}
each with $ z\in\bra{-\infty,\infty} $ and $ t > 0 $. Finally, the flow is driven by the oscillatory pressure gradient
\begin{equation}
	p_{0,z} = P_0 + P_1 e^{i t},\qquad r\in\left[0,1\right),\ z\in\left(-\infty,\infty\right),\ t > 0,
	\label{eqn:gov_p}
\end{equation}
with $ P_0 $ and $ P_1$ dimensionless constants.

We will now solve the boundary value problem~(\ref{eqn:gov_u}) -~(\ref{eqn:gov_cont}), and~(\ref{eqn:gov_noslip}) -~(\ref{eqn:gov_p}), hereafter termed [BVP], through constructing the $ O\bra{\epsilon} $ correction to the flow caused by the anastomosis.


\section{Analytic solution to the flow across an arterial anastomosis}
\label{sec:sol}

We begin by constructing the leading order approximation to [BVP]. Writing $ {w^{\pm} = w_0 + O\bra{\epsilon}} $, we obtain the leading order problem
\begin{eqnarray}
	\alpha^2 w_{0,t} - \frac{1}{r}\bra{r w_{0,r}}_r + P_0 + P_1 e^{i t} &=& 0,\qquad -\infty < z < \infty,\ 0 \leq r < 1,\ t > 0;\label{eqn:lo_w}\\
	w_{0,r}\bra{0,z,t} &=& 0,\qquad -\infty < z < \infty,\ t > 0;\label{eqn:lo_axisymw}\\
	w_0\bra{1,z,t} &=& 0,\qquad -\infty < z < \infty,\ t > 0.\label{eqn:lo_noslip}
\end{eqnarray}
The solution to the boundary value problem~(\ref{eqn:lo_w}) -~(\ref{eqn:lo_noslip}) is given by \citet{womersley1955method} for the case $ P_0 = 0 $, $ P_1 \neq 0 $, and, in the case $ P_0 \neq 0 $, $ P_1 = 0 $, by the solution to axisymmetric flow through a pipe. Thus, we have
\begin{equation}
	w_0\bra{r,z,t} = \frac{1}{4}P_0\bra{r^2 - 1} + \frac{i P_1}{\alpha^2}\bra{1 - \frac{\Jor}{\Jo}}e^{it},\label{eqn:lo_sol}
\end{equation}
on $ -\infty < z < \infty $, and $ 0 \leq r < 1 $, for $ t > 0 $, where $ J_k $ is the $ k^{th} $ order Bessel function of the first kind. It is important to note that the anastomosis has no effect on the leading order contribution to the flow. 

We now turn our attention to calculating the $ O\bra{\epsilon} $ correction to the flow caused by the anastomosis. Writing $ {u^{\pm} = u_1 + O\bra{\epsilon}} $, application of the solution~(\ref{eqn:lo_sol}) to the continuity equation~(\ref{eqn:gov_cont}), together with the no-penetration condition~(\ref{eqn:gov_nopen}), immediately gives that $ u_1 \equiv 0 $. Furthermore, following the form of the boundary conditions~(\ref{eqn:gov_noslip}) and~(\ref{eqn:gov_nopen}), we expand the velocity and pressure terms $ w $, and $ p $ as
\begin{equation}
	w = w_0\bra{r,z,t} +  \epsilon w_1\bra{r,z,t} + O\bra{\epsilon^2},\quad 	p = p_0\bra{z,t} +\epsilon p_1\bra{r,z,t} + O\bra{\epsilon^2},
	\label{eqn:fo_expansions}
\end{equation}
with $ w_0 $ and $ p_{0,z} $ given by~(\ref{eqn:lo_sol}), and~(\ref{eqn:gov_p}) respectively. On substitution of~(\ref{eqn:fo_expansions}) into [BVP] we obtain the boundary value problem at next order, namely
\begin{eqnarray}
	\alpha^2 w_{1,t} - \frac{1}{r}\bra{r w_{1,r}}_r + p_{1,z} &=& 0,\qquad -\infty < z < \infty,\ 0 \leq r < 1,\ t > 0;\label{eqn:no_w}\\
	p_{1,r} &=& 0,\qquad -\infty < z < \infty,\ 0 \leq r < 1,\ t > 0;\label{eqn:no_p}\\	
	w_{1,r}\bra{0,z,t} &=& 0,\qquad -\infty < z < \infty,\ t > 0;\label{eqn:no_axisymw}\\
	w_1\bra{1,z,t} \pm \ezb w_{0,r}\bra{1,z,t} &=& 0,\qquad -\infty < z < \infty,\ t > 0;\label{eqn:no_noslip}
\end{eqnarray}
where $ w_{0,r}\bra{1,z,t} $, calculated via~(\ref{eqn:lo_sol}), is given by
\begin{equation}
	w_{0,r}\bra{1,z,t} = \frac{1}{2}P_0 - \frac{i^{1/2} \Jbar P_1}{\alpha} e^{it},
	\label{eqn:no_noslip2}
\end{equation}
with the constant $ \Jbar = \Ji / \Jo $ having been introduced for algebraic convenience. The boundary value problem~(\ref{eqn:no_w}) -~(\ref{eqn:no_noslip}) admits the exact solution
\begin{eqnarray}
	\!\!\!\!\!\!w_1 &=& \pm\left[\frac{P_0}{2}\bra{1-2r^2} + \frac{i 2P_1\Jbar}{\alpha\bra{\alpha+2\Jbar i^{1/2}}}\bra{\Jbar -i^{3/2}\alpha \frac{\Jor}{\Jo}}e^{it}\right]\ezb,\label{eqn:no_sol_w}\\
	\!\!\!\!\!\!p_{1,z} &=& \mp\left[4P_0 - \frac{2\alpha P_1}{\alpha + 2\Jbar i^{1/2}} \Jbar^2 e^{it}\right]\ezb.\label{eqn:no_sol_p}
\end{eqnarray}
For completeness, the first non-zero component of the radial velocity is given, through expanding $ u^{\pm} = \epsilon u_2 + O\bra{\epsilon^2} $ and solving~(\ref{eqn:gov_cont}), together with~(\ref{eqn:gov_nopen}) and~(\ref{eqn:no_sol_w}), to give
\begin{equation}
	u_2 = \pm\left[\frac{P_0}{2\beta^2}\bra{r-r^3}z + \frac{i 2P_1 \Jbar}{\alpha\beta^2\bra{\alpha+2\Jbar i^{1/2}}}\bra{\Jbar r - \frac{\Jir}{\Jo}}z e^{it}\right]\ezb.
	\label{eqn:no_sol_u}
\end{equation}

Through substitution of the correction terms~(\ref{eqn:no_sol_w}),~(\ref{eqn:no_sol_p}), and~(\ref{eqn:no_sol_u}), together with the leading order solution~(\ref{eqn:lo_sol}), into the fluid velocity expansion~(\ref{eqn:gov_vel}) and taking the real part we obtain an analytical expression for the flow through a microvascular anastomosis. This modelled velocity profile enables the detailed and methodical investigation into the physical effects of introducing of a coupling device, as well as providing a first comparison with the case of sutured anastomosis.


\section{Results and analysis of the shear rate across the anastomosis}
\label{sec:disc}

In \S\ref{sec:sol} we constructed the $ O\bra{\epsilon} $ correction term to the Womersley flow profile, which arrises from modelling the case of a microvascular anastomosis through the inclusion of a small axisymmetric perturbation to the radial geometry. Where previous papers in this context (see \citet{wain2014}, \citet{wain2016}, \citet{wain2018}) have focussed on using CFD packages such as ANSYS-CFX to solve anastomoses problems with realistic geometries, the mathematical approach used here provides the basis for a detailed, rational exploration of the effect of conjoining small arteries with a radially symmetric coupling device. 

Throughout this section we will choose a realistic parameter set for the calculation and analysis of the results, namely the following. As the height of the bump due to a coupling device is difficult to accurately measure, in the present work we take the small parameter $ \epsilon = 0.04 / 1.25 $ to be equal to the size of a typical suture $\bra{0.04\mathrm{~mm}}$ relative to the typical radius of the vessels we are investigating $ \bra{1.25\mathrm{~mm}} $. The width parameter $ \beta = 0.3 $ is chosen arbitrarily for our calculations. The viscosity and density of blood give $ \mu^* = 0.0035 $~Pa~s, and $ \rho^* = 1060 $~kg~m$ ^{-3} $, and choosing the beat frequency $ \omega^* = 1.5 $~Hz then results in an approximate Womersley number $ \alpha \approx 0.84 $. Finally the pressure scales in~(\ref{eqn:gov_p}) are chosen to give a maximum dimensional velocity of $ 0.62$~ms$^{-1} $ (to match \citet{wain2018}) and to ensure no back flow, thus $ P_0 = P_1 \approx -660 $.

\begin{figure}
	\centering
	\includegraphics[width=\textwidth]{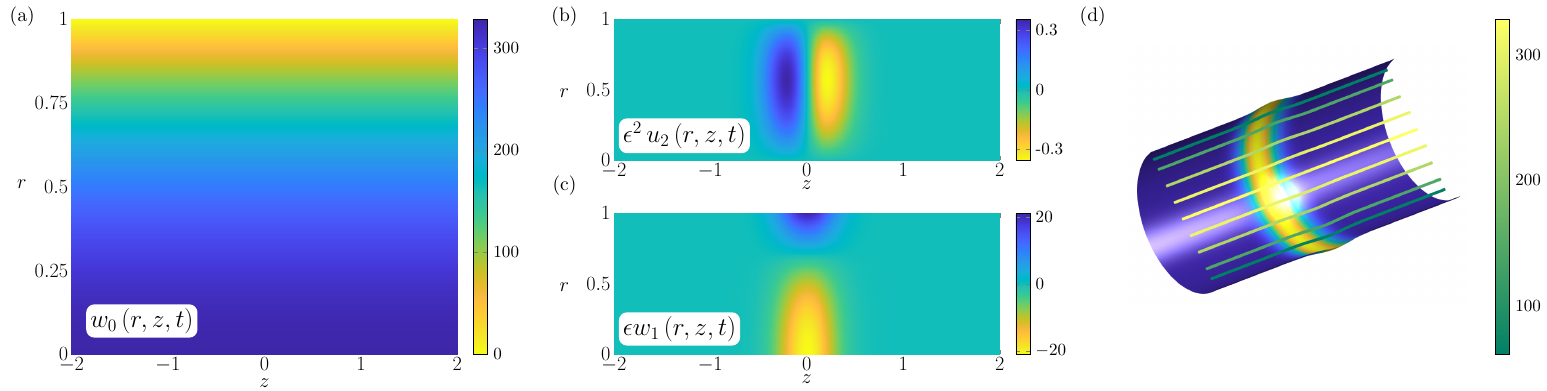}
	\caption{Velocity profile across the coupled vessel at time $ t = 0 $. In figure (a) we see the leading order contribution to the axial velocity, with the corrections to the velocity in the radial $ \bra{O\bra{\epsilon^2}} $  and axial $ \bra{O\bra{\epsilon}} $ directions in figures (b) and (c) respectively. Figure (d) shows a selection of streamlines through the not-to-scale coupled-vessel geometry and shear rate of figure \ref{fig:shr}c.}
	\label{fig:vel}
\end{figure}

In figure \ref{fig:vel} we see (a) the leading order and (c) the $ O\bra{\epsilon} $ correction to the axial velocity through a surgically coupled microvascular anastomosis, with (b) showing, for completeness, the $ O\bra{\epsilon^2} $ correction to the radial velocity, and (d) showing a selection of streamlines across the coupling. We see that, as expected, there is a small deformation to the streamlines, following the vessel wall, when compared to the pristine case. Far from the site of anastomosis the velocity profile relaxes to the pristine case.

\begin{figure}
	\centering
	\includegraphics[width=\textwidth]{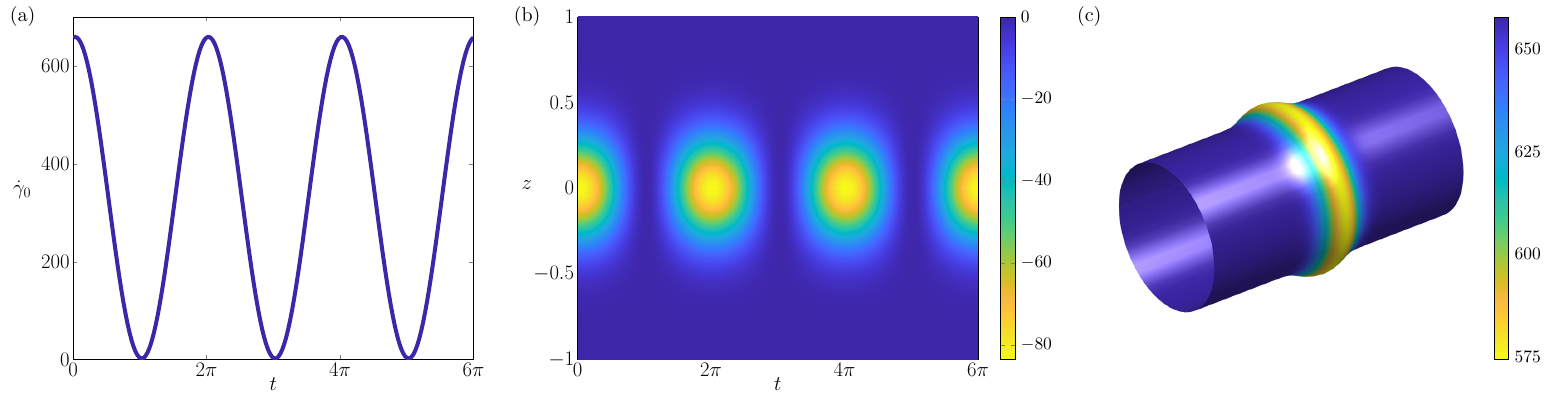}	
	\caption{Dimensionless shear rate across the vessel wall with time, calculated in~(\ref{eqn:shrRate}). In figure (a) we see the leading order spatially homogeneous contribution to the shear rate along the wall. The $ O\bra{\epsilon} $ correction to the shear rate is shown in (b), with the combined shear rate shown on the not-to-scale coupled-vessel geometry.}	
	\label{fig:shr}
\end{figure}

As discussed in \S\ref{sec:intro}, key to the rate of thrombus formation is both the shear rate and the spatial gradient of shear rate across the site of anastomosis. The dimensionless shear rate $ \dot{\gamma} $, non-dimensionalised with respect to the frequency parameter $ \omega^{*-1} $, is given, on substitution of the real part of the fluid velocity~(\ref{eqn:gov_vel}), as
\begin{equation}
	\dot{\gamma} = \left[\left(\frac{\partial w}{\partial r}\right)^2 + \epsilon^2\left(\frac{2u^2}{r^2} + \left(\pdiff{u}{r}\right)^2 + \left(\pdiff{w}{z}\right)^2 + 2\pdiff{u}{z}\pdiff{w}{r}\right) + \epsilon^4\left(\pdiff{u}{z}\right)^2\right]^{1/2}.
	\label{eqn:shrRate_full}
\end{equation}
Substitution of the asymptotic forms for $ u $ and $ w $ into~(\ref{eqn:shrRate_full}), provided $ \partial w_0 / \partial r \gg \epsilon$, yields the expansion
\begin{equation}
	\dot{\gamma} \sim \abs{\pdiff{w_{0}^{R}}{r}} + \epsilon\,\mathrm{sgn}\bra{\pdiff{w_{0}^{R}}{r}}\pdiff{w_{1}^{R}}{r},
	\label{eqn:shrRate}	
\end{equation}
where $ \mathrm{sgn} $ is the signum function. It is clear from~(\ref{eqn:shrRate}), together with~(\ref{eqn:no_sol_w}), that the correction to the shear rate in the case of an intruding bump differs from that of a protruding bump only in sign. Figure \ref{fig:shr} plots (a) the leading-order and (b) the $ O\bra{\epsilon} $ correction to the dimensionless shear rate with the parameter set defined above. Here we see that, in the vicinity of the anastomosis, the effect of the coupling device is a decrease in shear rate when compared to the pristine vessel, whereas the sutures effect a corresponding increase. The maximum change in shear rate compared to the pristine case is $ \sim 15\% $, and is found at the peak of the anastomosis. While this is much lower in the sutured case than the increase calculated by \citet{wain2018}, this is to be expected: in the case of a more realistic model of a suture the wall deformation would be concentrated over a smaller area, which is likely to effect a greater increase in shear rate than the initial model given here. It has also been noted by \citet{wain2016} that the angle of suture placement can have a significant effect on the shear stress when compared to those placed in-line with the direction of flow. 

Detailed examination of the asymptotic form of the shear rate~(\ref{eqn:shrRate}), together with the  axial velocity~(\ref{eqn:no_sol_w}), reveals that there is a linear relationship between the height of the wall deformation and the corresponding change to the shear rate, while the width parameter $ \beta $ has no effect on the magnitude of the $ O\bra{\epsilon} $ change in the shear rate, but controls the spatial gradient of the shear rate across the anastomosis through the spatial exponential factor in~(\ref{eqn:no_sol_w}).

We regards to the overarching aim of reducing the rates of thrombosis in microvascular anastomoses we can now draw the following conclusions. The results we have presented here agree with existing studies, showing a significant decrease in the shear rates due to the use of a surgical coupling device when compared to sutures, and thus coupling devices should be less likely to lead to thrombus growth. However, as discussed in \S\ref{sec:intro}, the magnitude of the shear rate does not tell the whole story: the spatial gradients of shear rate are also linked to platelet aggregation, and thus to increased rates of thrombosis. Through the detailed mathematical work presented here we have shown that the spatial gradients of shear rate, across the site of anastomosis, can be reduced by decreasing the gradient of the wall at the join, even if the maximum displacement on the wall remains the same (i.e. by increasing $ \beta $ in the model). 

\section{Discussion}

In this paper we have modelled the effect of microvascular anastomosis in small vessels of approximately $ 1 $~mm in radius on the flow driven by a pulsatile pressure gradient, with wall perturbations taking the form of both extruding and intruding axisymmetric deformations relating to the use of a coupling device as well as (a first analytical approximation to) the use of sutures. Through this work, the first analytical approach to these problems, we have been able to replicate the key findings of recent numerical work in this area, while providing new detailed insight into the ways in which new devices could be designed to minimise areas of potential biomedical impact. Key to this is the discovery that the width of the wall perturbation plays the significant role in the biologically important shear rate gradient, an insight which would not have been immediately apparent without the detailed analytical results presented here.

It is important to note that the asymptotic expansion for the shear rate~(\ref{eqn:shrRate}) is only valid providing $ \abs{\partial w_0/\partial r} \gg \epsilon $. When this is not the case for some point in time, i.e. in the case of flow reversal, a more detailed asymptotic examination is required. An interesting extension of this work would be to investigate the effect of a more accurate pressure gradient (for example, that derived from Doppler velocimetry in \citet{wain2018}). It is likely that in this case, some flow reversal would be apparent, requiring the more careful construction of the shear rate expansion.

A potential extension of the present work lies in the assumption that the fluid is Newtonian i.e. there is a linear relationship between the viscous stresses in the fluid and the strain rate. In regions of high shear rate this is likely to be a very good approximation, however when shear rates are more moderate a more physiologically accurate constitutive viscosity relationship may be required. In this case it is unlikely that a fully analytic solution would be forthcoming and as a result some amount of numerical approximation would be required.

The results from this model of microvascular anastomosis agree with the main findings from existing numerical studies in this area, while also providing new insight into potential ways for further reducing the thrombosis rates of surgical coupling devices through decreasing the change in spatial gradients of the shear rate. These findings support the growing clinical evidence in favour of using such devices in preference to the more traditional methods of suturing.

\section*{Acknowledgements}

M.T.G. is supported by the Engineering and Physical Sciences Research Council award EP/N021096/1. The author would like to thank David Smith and Thomas Montenegro-Johnson for helpful insights and discussions, Rich Wain and Justin Whitty for surgical and engineering related discussions, as well as the members of the Multi-scale Biology Study Group, University of Birmingham (12-15th December 2016) for valuable discussions regarding this subject.

The Multi-scale Biology Study Group, University of Birmingham (12-15th December 2016) was jointly funded by POEMS (Predictive modelling for healthcare technology through maths - EP/L001101/1) and MSB-Net (UK Multi-Scale Biology Network - BB/M025888/1).


\end{document}